\documentclass{article}

\PassOptionsToPackage{numbers, compress}{natbib}

\usepackage[preprint]{neurips_2022}

\usepackage[utf8]{inputenc} 
\usepackage[T1]{fontenc}    
\usepackage{hyperref}       
\usepackage{url}            
\usepackage{booktabs}       
\usepackage{amsfonts}       
\usepackage{nicefrac}       
\usepackage{microtype}      
\usepackage{xcolor}         
\usepackage{subcaption}
\usepackage{comment}
\usepackage{multirow}
\usepackage{tablefootnote}

\usepackage{amsmath}
\usepackage{graphicx}

\title{FEL: High Capacity Learning for Recommendation and Ranking via Federated Ensemble Learning}

\author{Meisam Hejazinia  Dzmitry Huba, Ilias Leontiadis, Kiwan Maeng, Mani Malek, \\ 
\textbf{Luca Melis, Ilya Mironov, Milad Nasr, Kaikai Wang, Carole-Jean Wu}\\\\ 
\textbf{Meta}
}

\begin{document}
\maketitle

\begin{abstract}
Federated learning (FL) has emerged as an effective approach to address consumer privacy needs. FL has been successfully applied to certain machine learning tasks, such as training smart keyboard models and keyword spotting. Despite FL’s initial success, many important deep learning use cases, such as ranking and recommendation tasks, have been limited from on-device learning. One of the key challenges faced by practical FL adoption for DL-based ranking and recommendation is the prohibitive resource requirements that cannot be satisfied by modern mobile systems.  We propose Federated Ensemble Learning (FEL) as a solution to tackle the large memory requirement of deep learning ranking and recommendation tasks. FEL enables large-scale ranking and recommendation model training on-device by simultaneously training multiple model versions on disjoint clusters of client devices. FEL integrates the trained sub-models via an over-arch layer into an ensemble model that is hosted on the server.  Our experiments demonstrate that FEL leads to 0.43--2.31\% model quality improvement over traditional on-device federated learning --- a significant improvement for ranking and recommendation system use cases. 


\end{abstract}

\section{Introduction}

Federated learning (FL) provides means to enhance data privacy for AI technologies. FL enables a large number of decentralized computing platforms at the edge to collaboratively train a shared machine learning (ML) model, while having raw data samples remain on-device. In the FL setting, the ML model is pushed to edge devices where data resides. In particular, local models are trained on edge devices, and updates to their parameters are shared with the central server using a secure aggregation protocol~\cite{secagg2017,huba2022papaya}. The server updates parameters of the global model and iterates by broadcasting them to the participating devices. FL has been deployed for a variety of machine learning tasks, such as smart keyboard \cite{abdulrahman2020survey}, personalized assistant services \cite{hao2020apple}, computer vision \cite{liu2020fedvision}, healthcare \cite{rieke2020future}, and ranking \cite{hartmann2019federated,paulik2021federated}.

Despite its growing profile, FL has seen limited adoption for ranking and recommendation tasks. 
The reasons are two-fold: (1) stringent accuracy requirement and (2) limited compute and memory capacity resources of client devices.
Recommendation systems are subject to strict accuracy requirements. A recent study from Baidu~\cite{zhao2020distributed} indicates that even a 0.1\% accuracy drop is considered unacceptable for its ranking and recommendation tasks.
However, due to the resource limitation of client devices, only small models with a constrained capacity (e.g., 20MB~\cite{huba2022papaya}) can be used.
The small model size requirement necessarily limits the achievable accuracy, as the learning capacity is much smaller than models traditionally used in the server setting, whose sizes are in the order of several GBs or~TBs~\cite{zhao2020distributed,lui:2021:capacity}.

While accuracy drop due to capacity constraints of the FL-trained model appears to be inevitable, it is not the case in the important setting of Label-only Privacy~\cite{ghazi2021deep,malek2021antipodes}. Indeed, the standard FL setup assumes that the entirety of users' data is private, and therefore both training and inference must be done on-device, severely limiting the model's size. On the other hand, if only the labels are deemed private, the inference can be performed server-side, potentially removing the model size pressure. Even if Label-only Privacy is assumed, however, it is unclear how a large server-side inference model can be trained using FL. 


One possible approach to train a large server-side model using FL with Label-only Privacy is to use split learning~\cite{vepakomma2018split, poirot2019split}. In the Label-only Privacy setting, split learning places only the last few layers of the model on client devices, while the rest of the model remains on the server. During training, the server and the client device exchange forward activations and backward gradients to jointly train the large model, without revealing the labels to the server.
However, split learning imposes several efficiency, scalability, and privacy challenges, as communication occurs at every forward/backward pass~\cite{vepakomma2018split} and backward gradients can leak label information~\cite{pasquini2021unleashing}.

To increase the effective learning capacity of FL in the Label-only Privacy case, we propose a new approach, \emph{Federated Ensemble Learning} (FEL). In FEL, multiple federated learning models, or \emph{leaf models}, are trained simultaneously on disjoint clusters of client devices. After being trained, the leaf models are combined together (ensembled) to form a larger model at the server.
The insight behind FEL is that if client device clusters are appropriately formed, leaf models can learn distinct, potentially complementary, knowledge from each cluster, which are later combined to enhance the overall prediction capability.
The idea stems from the logic that marketers have been using for hundreds of years: segmenting users into clusters allows fitting better products \cite{beane1987market} or models to each cluster.

In contrast with split learning, FEL combines communication efficiency with strong privacy guarantees.
Leaf models can be trained concurrently with each other when there is sufficient number of client devices. As each leaf model is trained using traditional FL, the privacy implication is similar to that of FL (Section~\ref{sec:privacy}).
FEL excels when the number of observations per user is small, but the number of users is high (e.g., in the order of billions), a setting which is common for recommendation and ranking tasks.

We have deployed FEL in the production environment of a large-scale ranking system. We show that the deployed recommendation task achieves 0.43\% precision gain compared to the vanilla FL baseline in the production environment, which is significant in the context of production ranking and recommender systems~\footnote{In similar applications, \cite{zhao2020distributed} mentions 0.1\% as significant and \cite{wang2017deep} considers 0.001 logloss, which is around 0.23\% in their context, as impactful.}.
We observe a similar gain of 1.55--2.31\% using the open-source datasets: one for Ads click prediction~\cite{taobao} and the other on image classification~\cite{liu2015deep}.
In addition, we show that FEL outperforms standard FL in the presence of differential privacy (DP) noise by 0.66--1.93\%.

\section{Background: Federated Learning and Privacy Assumptions}


{\bf Federated Learning (FL):}
FL trains a model collaboratively using clients' devices without clients' having to share their raw data with the server.
To train a model with FL, the central server first selects clients to participate from its client pool and broadcasts the model to selected clients. Then, each client trains the model using their own data, using their local computation resources. When the training is finished, each client sends back their trained model (or equivalently, the gradient)  to the server. Finally, the server collects the gradients and aggregates them (e.g., by taking a weighted average~\cite{mcmahan2017learning}) to update the server-side model. The process repeats until the model converges.


{\bf Scaling Federated Learning:}
The key constraint in using vanilla FL training for recommendation and ranking tasks is the limited model size it can support. In many federated recommendation and ranking tasks, the input space is large, e.g., over 1{,}000 features, and the data distribution is multi-modal. To achieve high accuracy in this setting, we have to increase the overall model capacity. However, FL can be applied only to sufficiently small models that can be trained on-device.

Prior work studied increasing the model capacity by leveraging client heterogeneity: training larger models on devices that are more powerful, while sending smaller models to less capable devices. The smaller model can be a subsampled model that is later aggregated to the supernet~\cite{fjord, expanding_reach, heterofl}, or a totally different model that later transfer its learned knowledge to the larger model with knowledge distillation~\cite{lin2020ensemble, fedmd}.
These approaches still limit the capacity of the model as the model size is capped by the most powerful client devices (e.g., high-end smartphones), which still cannot train GB-size models.

In the case of Label-only Privacy, it is possible to use an inference model that is too large to fit onto any single device. Still, it is unclear how such a large inference model can be trained when labels reside on the client devices.
Split learning splits the model and trains only the latter layers on the device, while training the rest on the server~\cite{vepakomma2018split, poirot2019split}.
Split learning, however, requires the device and the server to exchange intermediate activations, which can leak private label information~\cite{pasquini2021unleashing}. Also, many split learning approaches target cross-silo FL where only a handful clients participate~\cite{vepakomma2018split, poirot2019split} and it is unclear how these approaches can scale to cross-device FL with billions of devices, as each forward/backward pass of each client involves communicating with the server.

Other work used client models as the teacher to train a separate, larger server-side model~\cite{he2020group, cho2022heterogeneous}. These approaches, however, require both the client and the server to hold a representative public dataset to perform knowledge distillation~\cite{he2020group, cho2022heterogeneous}. In the real world, such a public dataset is unrealistic to assume in many scenarios. 



{\bf Privacy Assumptions of FEL:} FEL targets Label-only Privacy setting, where the input is public (i.e., accessible to the service provider) while the label is private. Several advertising, recommendation, and survey/analytics applications fall into this category~\cite{ghazi2021deep,malek2021antipodes,nalpasmore, pfeiffer2021masked}.

Many recommendation tasks use features that are public (vis-\`a-vis the recommender system). Public features include user attributes that are explicitly shared at sign-up time (e.g., age or gender)~\cite{taobao}, externally observable user behavior (e.g., public movie reviews)~\cite{movielens}, or item information that is provided by the item vendors~\cite{dlrm}. On the other hand, the labels of recommendation tasks may be considered private, as they often capture the users' conversion behavior (e.g., whether the user engaged with the recommended item by watching, clicking, buying, or signing up)~\cite{alibaba_fl, ghazi2021deep}.
It is also possible to use a mixture of public and private features.
FEL can be extended to incorporate private features (Section~\ref{sec:fel}).

While user labels must be kept private, i.e., unknown to the service provider, there can be \emph{opt-in users} who willingly agree to share their private label information to the service provider to improve the service quality. FEL does not require the presence of opt-in users; however, having some opt-in user population can simplify the training algorithm (Section~\ref{sec:fel}) and lower the privacy cost (Section~\ref{sec:privacy}).



To avoid statistical inference attacks targeting FL, 
differentially private (DP) noise is added either on device or during the aggregation step \cite{geyer2017,truex2019,wei2020federated}. There are various notions of privacy, but we focus on user-level differential privacy, in which the trained model weights are similarly distributed with or without a particular user~\cite{mcmahan2017learning}.
We assume an honest-but-curious provider for training FL, which uses either hardware-based encryption in a trusted enclave, or software-based encryption via multiparty computation for FL aggregation \cite{li2020privacy,mondal2021flatee}.

\section{Proposed Design: Federated Ensemble Learning (FEL)}
\label{sec:fel}

Rather than training one model across all users, FEL proposes to train a distinct leaf model per user cluster and later aggregate the leaf models on the server to obtain a larger-capacity model.
Ensemble methods similar in spirit have shown promising potential with classical machine learning algorithms, e.g., in the form of AdaBoost, random forest, and XGBoost \cite{le2021fedxgboost}.
We explored different variations in how the clients are clustered and how the leaf models are aggregated.


\subsection{Federated Ensemble Learning} 

\begin{figure}
\centering
\includegraphics[width=1\textwidth]{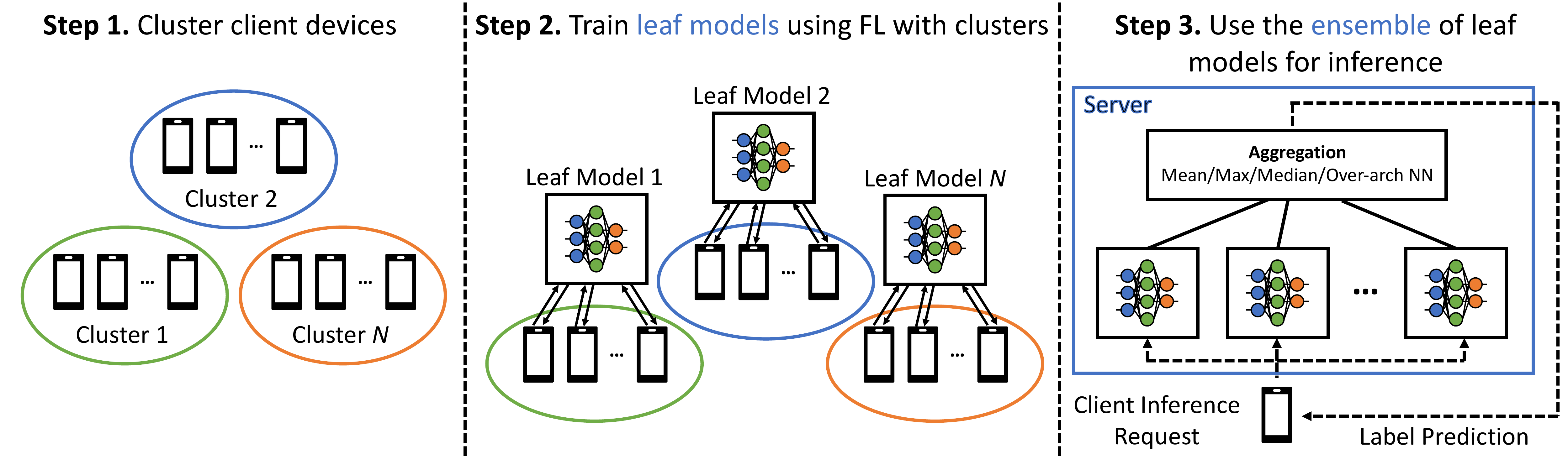}
\caption{\label{fig:FEL_architecture.png} Federated Ensemble Learning architectures}
\end{figure}

Figure~\ref{fig:FEL_architecture.png} illustrates the proposed design of FEL. First, we cluster the clients into a desired number of clusters (Figure~\ref{fig:FEL_architecture.png}, Step 1). Then, we train a leaf model for each cluster using traditional FL (Figure~\ref{fig:FEL_architecture.png}, Step 2). After training, the server uses the ensemble of the leaf models for inference requests (Figure~\ref{fig:FEL_architecture.png}, Step 3).
On each inference request, the server passes the public input features through all the leaf models. Then, the output of each leaf model and/or the output of the last hidden layer of each leaf model is passed to the \emph{ensemble aggregation} layer, which outputs the final prediction.
Below, we lay out how FEL forms clusters and implements ensemble aggregation, and how FEL can be extended to incorporate private features.

{\bf Forming Clusters}
Client clusters can be formed based on distinctive characteristics of the users, e.g., a user's age, location, or past preferences. Clusters can be also obtained by simple hashing, or through popular clustering approaches such as k-means or Gaussian mixture methods. Marketers have been forming clusters of clients to target each cluster more effectively, and those well-studied clustering methods can be adopted~\cite{beane1987market}.

{\bf Rudimentary Ensemble Aggregation:}
A simplest way to implement ensemble aggregation is to collect the prediction output of each leaf model and perform typical aggregation methods, such as \textit{mean}, \textit{median}, or \textit{max}, to generate the final prediction. This approach is similar to bagging technique leveraged in random forest \cite{prasad2006newer}.

{\bf Neural Network (NN)-based Ensemble Aggregation:}
NN-based ensemble aggregation uses a separate neural network that takes in the prediction and the output of the last hidden layer of all leaf models as input to generate the final prediction.
We call this additional neural network model \textit{the over-arch model}.
The over-arch model is trained after all the leaf models are trained. The over-arch model can be trained in several different ways. In the presence of opt-in users, the server can simply use them to train the over-arch model. Otherwise, the server can again use FL to train the over-arch model on each client, where the server sends the output of the leaf models to the client, and the client uses it as an input to train the over-arch model locally.


{\bf Extending to Private Features}
FEL can be extended to support private features only known to the clients. This can be done by (privately) training a separate leaf model (private leaf model) that only takes in private client-side features. The output of the private leaf model is used as an input to the ensemble aggregation layer along with other leaf models.
When the private leaf model is used, part of the inference must happen on-device instead of entirely on the server. Specifically, the server sends outputs of the leaf models to each client, which ensembles them with its private leaf model output on-device for prediction. By performing the ensemble on-device, the input/output of the private leaf model is never exposed to the server. The private leaf model is trained with conventional FL as well.


\subsection{Privacy Analysis for Federated Ensemble Learning}
\label{sec:privacy}

Differential privacy for federated learning bounds how much model parameters change for two datasets with only one different user \cite{mcmahan2017learning}. Formally we define:

\textbf{Definition 1} (Renyi Differential Privacy (RDP)). A \textit{randomized mechanism $\mathrm{M}$ with domain $\mathrm{D}$ is $(\alpha, \epsilon)$-RDP with order $\alpha \in (1,\infty)$ iff for any two neighboring datasets $D, D'\in \mathrm{D}$:}

\begin{align*}
  \mathrm{D}_\alpha(\mathrm{M}(D)||\mathrm{M}(D')):=\frac{1}{\alpha - 1}\log_{\delta \sim \mathrm{M}(D')}E\left[(\frac{\mathrm{M}(D)}{\mathrm{M}(D')})^\alpha\right]\leq \epsilon
\end{align*}

FEL is a multi-step process. To analyze the privacy bounds of a multi-step process, a common approach in differential privacy is to evaluate each process individually, then calculate the overall privacy bounds by composing all of the steps. In particular, we focus on two main composition theorems in differential privacy, sequential and parallel composition \cite{dwork2009differential}. Formally we define:

\textbf{Theorem 1} (Sequential Composition). \textit{Let there be n RDP-mechanisms $\mathrm{M}_i$ with $(\alpha, \epsilon_i)$-RDP when being computed on a dataset $D$ of the input domain $\mathrm{D}$. Then, the composition of n mechanisms $\mathrm{M}(\mathrm{M}_1(D),\dots,\mathrm{M}_n(D))$ is $(\alpha, \sum^n_{i=1} \epsilon_i)$-RDP}

\textbf{Theorem 2} (Parallel Composition). \textit{Let there be n RDP-mechanisms $\mathrm{M}_i$ with $(\alpha, \epsilon_i)$-RDP when being computed on disjoint subset $\mathrm{D}_i$ of the input domain $\mathrm{D}$. Then, the composition of n mechanisms $\mathrm{M}(\mathrm{M}_1(D),\dots,\mathrm{M}_n(D))$ is $(\alpha, \max^n_{i=1} \epsilon_i)$-RDP}

Sequential composition considers the case where a task uses the same users (even if different steps use different parts of the user’s data) in different steps of an algorithm. For example, if the algorithm has four steps each with a privacy cost $e$ and uses the same users in all the steps, the total privacy cost of the algorithm would become $4e$.
Parallel composition considers a case where each step is applied to different users. From the earlier example, if four disjoint set of users were used for the for steps, the overall privacy cost would be $\max(e_1,e_2,e_3,e_4)$, where $e_k$ is the privacy cost of the $k$-th step. It should be noted that the compositions are analyzed using RDP. The additivity is not exactly linear in $(\epsilon, \delta)$-DP, and while possible, $(\epsilon, \delta)$-DP cannot yield as tight bounds as RDP.

When analyzing the privacy cost of training the leaf models, the parallel composition theorem applies as each leaf model is trained with a disjoint set of users. If training a leaf model with cluster $i \in \{1,\dots,N\}$ has a privacy cost $e_i$, the privacy cost of the entire leaf model is $e_\mathrm{leafs} = \max(e_1,\dots, e_N)$.

The privacy cost of the ensemble aggregation layer $e_\mathrm{agg}$ depends on the aggregation method that is used (mean, max, median, or NN-based).
When using the mean, max, and median ensemble aggregation, $e_\mathrm{agg}=0$ as both theorems show that privacy cost increases only when the additional step (ensemble aggregation) uses user data.
In this case, the total privacy cost $e_\mathrm{tot}$ is simply $e_\mathrm{leafs}$.

When NN-based approach is used, however, user data is used to train the over-arch NN layer and $e_\mathrm{agg} > 0$. Depending on exactly how the over-arch NN layer is trained, $e_\mathrm{tot}$ can be calculated in the following way.
First, if the over-arch NN layer is trained with the users that trained the leaf models, sequential composition theorem applies ($e_\mathrm{tot} = e_\mathrm{agg} + e_\mathrm{leafs}$).
Second, if the over-arch NN layer is trained with a completely different set of users, the parallel composition theorem applies ($e_\mathrm{tot} = \max(e_\mathrm{agg}, e_\mathrm{leafs})$).
Finally, if the over-arch NN layer is trained with opt-in users, $e_\mathrm{agg}=0$ because no private data is used, and $e_\mathrm{tot}=e_\mathrm{leafs}$.
In all cases, the privacy cost of FEL does not significantly deteriorate over the vanilla FL (which is similar to $e_\mathrm{leafs}$).





\section{Experimental Methodology}

\subsection{Datasets}
We use three datasets for the purpose of this study. To study recommendation and ranking tasks, we used a production dataset and an open-source, Taobao's Click-Through-Rate (CTR) prediction dataset~\cite{li2021novel}.
To study the effect of FEL on non-recommendation use-cases, we additionally studied the LEAF CelebA Smile Prediction dataset~\cite{liu2015deep}.

{\bf Production Dataset:}
Production dataset is an internal dataset that captures whether a user installs a mobile application after being shown a relevant advertisement item. A few hundred features are used as an input (the exact number cannot be disclosed) to predict a binary label (install/not-install). All the input features are public. For training, we use advertisement data from a random sample of 35 million users over a period of one month. Randomly selected 15 million users from the following week were used for testing.  


{\bf Taobao CTR Dataset:}
Taobao dataset contains 26 million interactions (click/non-click when an Ad was shown) between 1.14 million users and 847 thousand items across an 8-day period. 
The dataset uses 9 user features (e.g., gender or occupation), 6 item features (e.g., price or brand), and two contextual features (e.g., the day of week), which we assume to be all public to the service provider. 

In the Taobao CTR dataset, 16 out of the 17 features are sparse, with a categorical value encoding instead of a continuous, floating point value.
While server-based recommendation models use large embedding tables to convert these sparse features into a floating point embedding~\cite{din, dlrm, wideanddeep}, training such embedding tables on device is complicated because of the large memory capacity requirement (e.g., in the order of GB to TB~\cite{zhao2020distributed,acun:2021:understandingtraining,wilkening:2021:recssd,lui:2021:capacity}) and can leak private information more easily through gradients~\cite{alibaba_fl}. 
Thus, we assume an architecture where embedding tables are pre-trained with opt-in users and are hosted on the server, while the rest of the model is trained with FEL using sparse features translated through the pre-trained tables. We randomly selected 10\% of the users as opt-in.

Note that our setup cannot achieve the accuracy that can be reached when we fully train the embedding tables, as we pre-train the embedding table and fix their weight during FL.
However, our setup represents a practical FL setup where training embedding tables on-device is prohibitive, due to client resource limitations~\cite{nguyen2021federated} and privacy concerns~\cite{alibaba_fl}.

{\bf CelebA Smile Prediction Dataset:}
While FEL is originally designed for recommendation and ranking tasks, we study its generality to non-recommendation models with CelebFaces Attributes Dataset (CelebA)~\cite{liu2015deep}. CelebA consists of $200,288$ images belonging to $9,343$ unique celebrities. Each image has 40 binary facial attribute annotations (e.g., bald, long hair, attractive, etc) and covers large pose variations and backgrounds.
We defined distinguishing between smiling/non-smiling images as our target task.

\subsection{Model Architectures}

{\bf Production/Taobao Dataset:}
For recommendation datasets (production/Taobao CTR), we use a model that consists of 3 fully-connected hidden layers. The number of units at each hidden layer is decreasing exponentially with a parameter K. For instance, if $K=4$ and the input layer has 512 features, our neural network would have  $[512, 128, 32, 8, 1]$ neurons. For each dataset, we tune K to obtain a resulting model of approximately $10$MB. By doing so, it allows us to train a neural network even on older, low-tier devices with more limited memory capacity. ReLu is used as an activation function after each layer apart from the last one, where Sigmoid and binary cross-entropy was used.

For both datasets, we use synchronous FL with FedAvg~\cite{mcmahan2017learning}. We used the following hyperparameters for the Taobao dataset from an extensive hyperparameter search: client batch size of 32, 5 local epochs, 4096 clients per round, and a learning rate of 0.579 with SGD. Clients are selected at random and each only participates once (1 global epoch). The production dataset used similar hyperparameters.

For Taobao dataset's server-side pre-trained embedding table, we use an embedding dimension of 32, and train it with the 10\% opt-in users for 1 epoch using AdaGrad optimizer with learning rate of 0.01.

{\bf CelebA Dataset:}
For CelebA, we follow the setup of prior work~\cite{nguyen2021federated} and use a four layer CNN with dropout rate of 0.1, stride of 1, and padding of 2. We preprocess all images in train/validation/test sets; each image is resized and cropped to 32$\times$32 pixels, then normalized by 0.5 mean and 0.5 standard deviation. We use asynchronous FL with a client batch size of 32 samples, 1 local epoch, 30 global epochs, and a learning rate of 0.899 with SGD.

\subsection{FL Baseline and FEL}
Both the FL baseline and the FEL leaf models used the same set of hyperparameters. 
The FL baseline is trained using all the available client data. In FEL, the client data is clustered, and one leaf model is trained for each cluster. We vary the number of clusters from 3--10 and evaluate different clustering methods.
When training the over-arch NN layer, a small subset of opt-in users is used.

\section{Evaluation Results and Analysis}

\begin{table}
\small
\centering
\begin{tabular}{|l||l|l|c|}
\hline
Dataset & Config & Feature & \# clusters \\ \hline\hline
\multirow{4}{*}{Production} & Clustering 1 & Age & 5\\
 & Clustering 2 & App & 5\\
 & Clustering 3 & Location & 4\\
 & Clustering 4 & Click ratio & 10\\\hline
\multirow{3}{*}{Taobao~\cite{taobao}} & Clustering 1 & Age & 7\\
 & Clustering 2 & Consumption & 4\\
 & Clustering 3 & City level & 5\\\hline
\multirow{3}{*}{CelebA~\cite{liu2015deep}} & Clustering 1 & \# Attributes & 3\\
 & Clustering 2 & K-means & 3\\
 & Clustering 3 & K-means & 5\\\hline
\end{tabular}\\
\vspace{0.25cm}
\caption{\label{tab:clusterconfig} Explanation of different cluster methods in Figure~\ref{fig:eval0} (right).}
\end{table}

Our evaluation aims to answer the following questions:

\begin{itemize}
    \item Can FEL improve the model prediction quality over vanilla FL? [Section~\ref{sec:prediction-quality}]
    \item How do different ensemble aggregation methods affect the model accuracy? [Section~\ref{sec:prediction-quality}]
    \item How do different clustering methods affect the model accuracy? [Section~\ref{sec:prediction-quality-clustering}]
    \item How does FEL affect privacy compared to vanilla FL? [Section~\ref{sec:privacy-eval}]
\end{itemize}

\subsection{Prediction Quality Improvement of FEL}
\label{sec:prediction-quality}
Overall, FEL achieves \textbf{0.43\%} and \textbf{2.31\%} prediction quality improvement over vanilla FL for production and Taobao datasets, respectively -- a significant improvement for ranking and recommendation system use cases\footnote{\cite{zhao2020distributed} mentioned 0.1\% model quality improvement as significant and \cite{wang2017deep} considered 0.23\% as impactful in similar recommendation and ranking use-cases.}.
For non-recommendation tasks (CelebA), FEL shows similar improvement of~\textbf{1.55\%}, indicating that FEL can be generalized to non-recommendation use-cases as well.
Table~\ref{tab:result1} summarizes the resulting prediction quality improvement of FEL compared to the baseline FL. Following common practice of each dataset, we used accuracy for CelebA~\cite{nguyen2021federated} and ROC-AUC (AUC) for Taobao~\cite{taobao}. We used normalized entropy for the production dataset, which we cannot disclose and only show the relative improvement.
For different ensemble aggregation methods, we vary the clustering methods and report the best-accuracy results.

\begin{table}
\centering
\begin{tabular}{|l|cccc|}
\hline
 & Production & Taobao~\cite{taobao} AUC & CelebA~\cite{liu2015deep} accuracy & Geomean \\ \hline\hline

Baseline FL & - & 0.5418~\tablefootnote{Taobao's baseline AUC is 0.26\% less than the baseline FL result presented at~\cite{alibaba_fl}, potentially due to simpler model architecture and freezed pre-trained embedding tables.} & 90.75 & - \\\hline
FEL (Mean Best) & (+0.27\%) & 0.5522 (+1.92\%) & 91.68 (+1.02\%) & (+1.07\%)\\
FEL (Median Best) & (+0.29\%) & 0.5459 (+0.74\%) & 91.35 (+0.66\%) & (+0.56\%)\\
FEL (Max Best) & (-0.06\%) & 0.5418 (-0.1\%) & 91.46 (+0.78\%) & (+0.21\%)\\
FEL (NN-based Best) & \textbf{(+0.43\%)} & \textbf{0.5544 (+2.31\%)} & \textbf{92.16 (+1.55\%)} & \textbf{(+1.43\%)}\\\hline
\end{tabular}\\
\vspace{0.25cm}
\caption{\label{tab:result1} FEL's prediction accuracy improvement over the baseline FL for different datasets. Following common practice of each dataset, Taobao uses AUC and CelebA uses accuracy as their metric. Production data's baseline accuracy is not disclosed.}
\end{table}

Among the different ensemble aggregation methods, adding an over-arch NN layer provided the best prediction quality improvement, followed by mean and median. Max only showed improvement in CelebA and did not show benefit in recommendation use-cases.

\subsection{Prediction Quality Improvement of Different Clustering Methods}
\label{sec:prediction-quality-clustering}

\begin{figure}
\centering
    \begin{subfigure}[b]{0.45\textwidth}
        \centering
        \includegraphics[width=\textwidth]{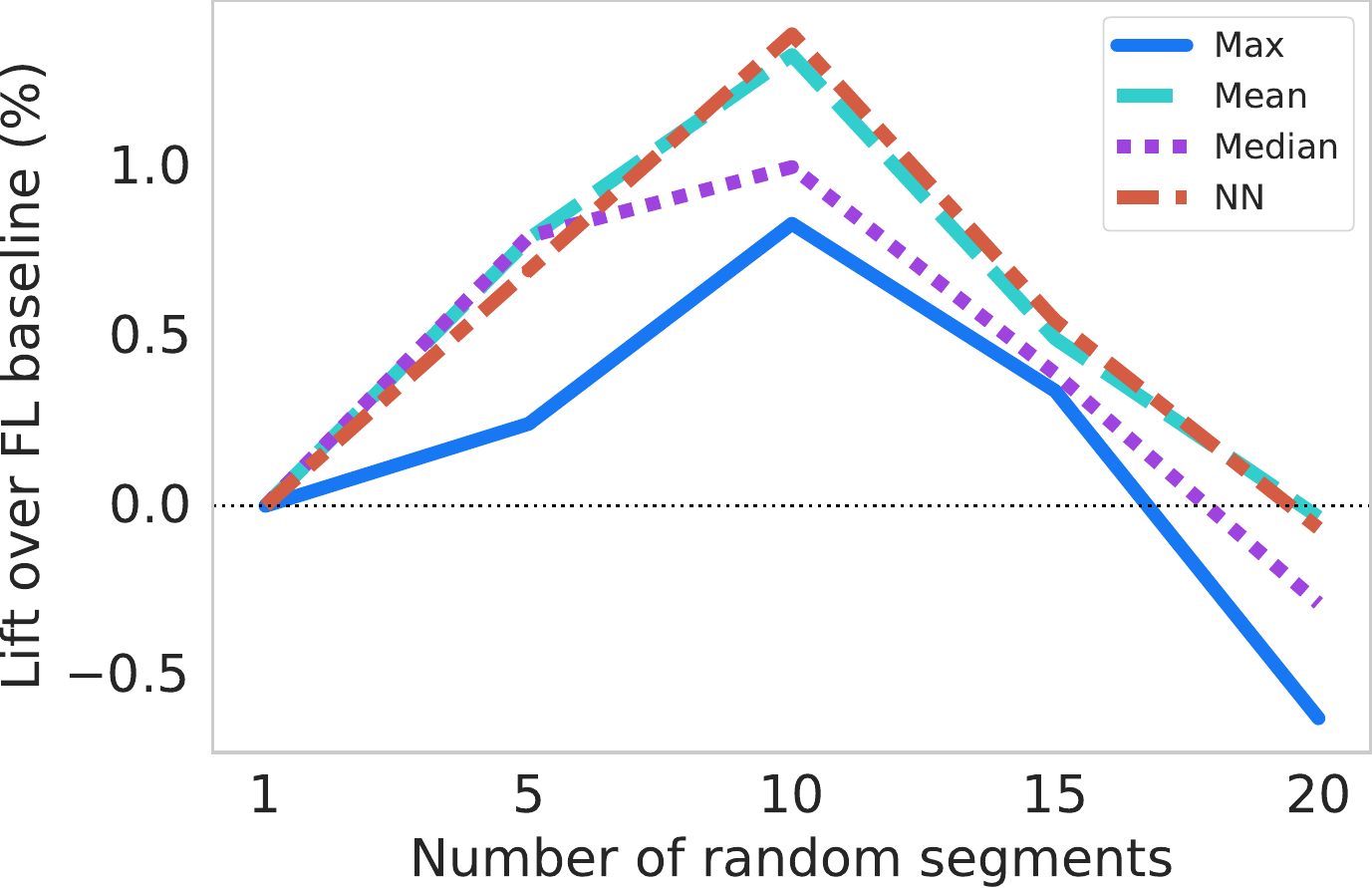}
        \label{fig:eval0_0}
     \end{subfigure}
     \begin{subfigure}[b]{0.49\textwidth}
        \centering
        \includegraphics[width=\textwidth]{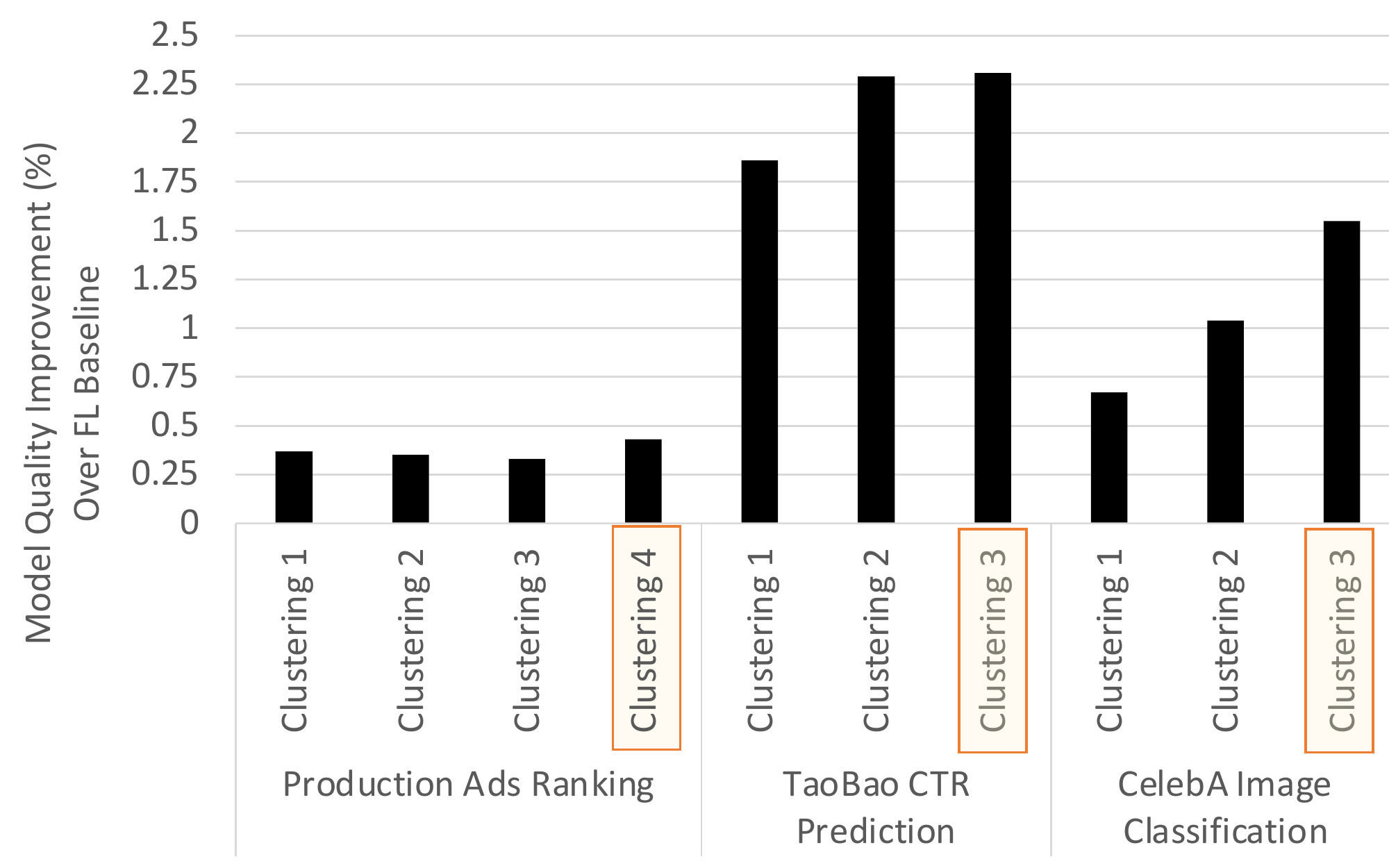}
        \label{fig:eval0_1}
     \end{subfigure}
     \caption{Accuracy improvement for different number of clusters (segments) for each ensemble aggregation method (left), and different clustering methods when using over-arch NN layer (right). Different clustering methods are explain in Table~\ref{tab:clusterconfig}}
    \label{fig:eval0}
\end{figure}

{\bf Effects of the Number of Clusters:}
To understand the effect of the number of client clusters in the final model quality improvement, we varied the number of clusters in the Taobao dataset while using random clustering.
Figure~\ref{fig:eval0} (left) summarizes the result. There is an optimal setting for the number of clusters used in FEL. Going beyond the optimal setting for the number of clusters results in worse model accuracy. When the number of clusters is too small, the final model capacity is limited as there are not enough leaf models to ensemble. If the number of clusters is too large, each leaf model cannot learn enough information as the clients in each cluster are too few. The optimal number of clusters depends on the number of available devices that participate within each cluster and, here, the number of partitions can be treated as a hyperparameter~\cite{kim:micro21}. 

{\bf Effects of Features Used in Clustering:}
We also varied the clustering methods for each dataset and observed their effect on the final model accuracy. We explored different clustering methods for different datasets and presented the best performing methods. Table~\ref{tab:clusterconfig} summarizes the clustering methods that we explored. Here, we show the result for the best performing over-arch NN-based ensemble aggregation for brevity. For the production dataset, we used user age, the app category where the ad was displayed, location (larger geographic regions), and previous click ratio of the users to cluster the users. For Taobao dataset, we used user age, city level, and consumption level. For CelebA, we clustered the 40 binary attributes of each user using K-means clustering or simply used the number of present attributes.

Figure~\ref{fig:eval0} (right) shows that clustering can affect the final model accuracy significantly. For the production dataset, clustering using the click ratio (Clustering 4) showed the best accuracy. For Taobao, clustering with city level showed the best accuracy (Clustering 3). For CelebA, using K-means clustering was the best (Clustering 3). The results show that clustering methods as well as the number of clusters are two important hyperparameters of FEL.

\subsection{Evaluation Results with Differential Privacy}
\label{sec:privacy-eval}

\begin{table}
\centering
\begin{tabular}{|lc|cccc|}
\hline
Config & $\epsilon$ & Mean & Median & Max & Over-Arch NN  \\ \hline\hline
\multirow{3}{*}{Clustering 1} & $\infty$ & +1.27\% & -0.23\% & -0.10\% & \textbf{+1.86\%} \\
 & 3.78 & +0.63\% & -0.14\% & +0.11\% & \textbf{+0.66\%} \\
 & 1.56 & +0.32\% & -0.23\% & +0.34\% & \textbf{+0.68\%} \\ \hline

\multirow{3}{*}{Clustering 2} & $\infty$ & +1.92\% & -0.46\% & -1.64\% & \textbf{+2.29\%} \\
 & 3.78 & +0.75\% & -0.21\% & +0.03\% & \textbf{+1.38\%} \\
 & 1.56 & +0.69\% & -0.11\% & +0.74\% & \textbf{+0.76\%} \\ \hline
 
\multirow{3}{*}{Clustering 3} & $\infty$ & +1.86\% & +0.74\% & -2.07\% & \textbf{+2.31\%} \\
 & 3.78 & +1.49\% & -0.09\% & +1.03\% & \textbf{+1.93\%} \\
 & 1.56 & +0.71\% & -0.37\% & +1.26\% & \textbf{+1.02\%} \\ \hline
 
\end{tabular}\\
\vspace{0.25cm}
\caption{\label{tab:FEAdsResultsDP} Taobao dataset with DP. Percentage of FEL's accuracy improvement over the FL baseline with the same level of DP noise is shown. Table~\ref{tab:clusterconfig} explains the clustering configurations.}
\end{table}

Table~\ref{tab:FEAdsResultsDP} shows the accuracy improvement of FEL compared to vanilla FL for two different levels of DP noise, along with the case of no DP noise ($\epsilon=\infty$). We assume the over-arch NN layer was trained with opt-in data and no DP noise added when training the over-arch NN layer.
Table~\ref{tab:FEAdsResultsDP} shows that even when DP noise is added, FEL shows meaningful accuracy improvement over vanilla FL. 
Again, we observe that the over-arch NN layer and mean aggregations still provide the most significant gains. 
However, smaller $\epsilon$ leads to reduced accuracy gain, possibly due to larger injected noise. 
%
Another interesting observation is that the max ensemble aggregation improves the accuracy when DP noise is added, unlike the no DP noise case where it did not show any improvement. One possible reason is because DP noise mitigates the effects of outliers in training.

\section{Related Work}
This study resides in the intersection of four areas of study: ensemble distillation, boosted federated learning, local ensemble learning, and ensemble aggregation.

\textbf{Ensemble Distillation.} Lin et al.~\cite{lin2020ensemble} propose FedDF that uses unlabeled data generated by a generative model to aggregate knowledge from all heterogeneous client models, rather than leveraging FedAVG. This model uses average logit and fusion to share learning between heterogeneous models. Gong et al.~\cite{gong2022preserving} focus on communication efficiency and privacy guarantee with one-shot offline knowledge distillation. This proposal keeps the local training asynchronous and independent, and then aggregates the local predictions on unlabeled cross-domain public data. Sui et al.~\cite{sui2020feded} explore a knowledge distillation strategy which uses the uploaded predictions of ensemble local models to train the central model without requiring uploading local parameters. This approach only uses predicted labels on a small dataset to learn a student model from an ensemble of multiple local teacher models. 

\textbf{Boosted Federated Learning.} Boosting and bagging are two prominent approaches for model ensemble learning. Li et al.~\cite{li2020practical} distribute data samples with the same features among multiple parties, relaxing privacy concerns. In their approach, each party boosts a number of trees by exploiting similarity information based on locality-sensitive hashing. In Hamer et al.~\cite{hamer2020fedboost} work, an ensemble of pre-trained based predictors is trained via federated learning, thus saving on communication costs. Luo et al.~\cite{luo2021research} suggest gradient boosting decision tree (GBDT) method, which takes the average gradient of similar samples and its own gradient as a new gradient to improve the accuracy of the local model.

\textbf{Local Ensemble Learning.} Shi et al.~\cite{shi2021fed} propose FedEnsemble which uses random permutations to update a group of $K$ models, and then obtains predictions through model averaging, instead of aggregating local models to update a single global model. Similarly, Majeed et al.~\cite{majeed2020blockchain} suggest an ensemble learning FL regime in which five base FL models are trained using the same local datasets, and ensemble using simple majority voting rule. Attota et al.~\cite{attota2021ensemble} propose MV-FLID, which is a multi-view ensemble learning which helps in maximizing the learning efficiency of different classes of attacks for intrusion detection tasks. 

\textbf{Ensemble Aggregation.} Chen et al.~\cite{chen2020fedbe} suggest FedBE, which takes a Bayesian inference perspective by sampling higher-quality global models and combining them via Bayesian ensemble for robust aggregation. Guha et al.~\cite{guha2019one} propose one-shot learning, where a central server learns a global model over a network of federated devices in a single round of communication. Liu et al.~\cite{liu2020privacy} suggest FedGRU, for business-to-business (B2B) setting, which uses both secure parameter aggregation and cluster ensembles to scale.  Bian et al.~\cite{bian2021fedseal} extend federated ensembles to the context of semi-supervised learning, leveraging self-ensembles, to enable clients to label their own data. Orhobor et al.~\cite{orhobor2020federated} suggest assigning users into pre-specified bins and train different regressors on each bin, which are later ensembled for precision medicine. 

Although these methods have their own merits, they do not address the problem of the recommender and ranking systems use cases, in which each user has only a small number of examples, and require user-level privacy guarantee. As a result, none of these studies leverages the variation across users and diversity of behavior in their proposals. Our approach clusters the users and trains different models on different users, leveraging a large user base in recommender and ranking systems. Furthermore, our user partitioning, feature generation by last hidden layer, and training over-arch model provides extra gains both in terms of precision and privacy budget.

\section{Conclusion}
While Federated Learning is gaining traction for select applications, it cannot be directly adopted to ranking and recommendation tasks that require large model size.
We introduce Federated Ensemble Learning (FEL), which increases the learning capacity of FL models.
FEL clusters  client population and trains a leaf model per each cluster, which are later ensembled to form a larger inference model.
FEL can be trained efficiently without introducing significant privacy concerns and can improve the prediction accuracy meaningfully compared to vanilla FL.
FEL enables FL for demanding ranking and recommendation tasks. 

As future work we plan to evaluate FEL on different model architectures, such as transformer, LSTM, RNN, or CNN and on other tasks such as speech recognition, reinforcement learning.
FL-based design must also consider the inter-dependence of data and system heterogeneity, as observed in real-world, large-scale federated recommendation learning~\cite{maeng:recsys2022}.
Furthermore, we plan to integrate unsupervised clustering approaches, so that the segmentation can happen automatically to optimise FEL's performance.
%
Finally, to minimise the cost of managing a number of leaf models, we plan to study automated ways to assess the quality of each leaf model  to pinpoint under-represented clusters and seek possible mitigation such as dynamic clustering and leaf retraining.

\section*{Acknowledgement}
We would like to thank Milan Shen, Will Bullock, Hung Duong, and Kim Hazelwood for supporting the work.

\bibliographystyle{plainnat}
\bibliography{sample}

\end{document}